\documentclass{article}
\usepackage[utf8]{inputenc}
\usepackage{authblk}
\usepackage{setspace}
\usepackage{mathdots}
\usepackage[margin=1.25in]{geometry}
\usepackage{graphicx}
\usepackage{subcaption}
\usepackage{amsmath}
\usepackage{amssymb}
\usepackage[style=nejm, 
citestyle=numeric-comp,
sorting=none]{biblatex}
\title{Attosecond Waveform Synthesis through Echo-enabled Harmonic Generation Free-electron Lasers}

\author[1,2]{Lanpeng Ni}
\author[3]{Junhao Liu}
\author[4*]{Zheng Qi}
\author[4*]{Chao Feng}

\affil[1]{Shanghai Institute of Applied Physics, Chinese Academy of Sciences, Shanghai 201800, China.}
\affil[2]{University of the Chinese Academy of Sciences, Beijing 100049, China.}
\affil[3]{ShanghaiTech University, Shanghai 201210, China.}
\affil[4]{Shanghai Advanced Research Institute, Chinese Academy of Sciences, Shanghai 201210, China.}
\affil[*]{Address correspondence to: qiz@sari.cn, fengc@sari.ac.cn}

\date{}

\begin{document}

\maketitle
%%%%%% Abstract %%%%%%
% \begin{abstract}
% %The abstract should be a single paragraph written in plain language that a general reader can understand. Do not include citations, figures, tables, or undefined abbreviations in the abstract. Any abbreviations that appear in the title should be defined in the abstract. The length should be 200 words and not exceed 250 words, to include: 
% % \begin{itemize}
% %     \item An opening sentence that states the question/problem addressed by the research AND
% %     \item Enough background content to give context to the study AND
% %     \item A brief statement of primary results AND
% %     \item A short concluding sentence.
% % \end{itemize} 
\begin{abstract}
Attosecond pulse trains (APTs) are indispensable for probing electron dynamics at their intrinsic timescales. High harmonic generation (HHG) has long been a successful and widely used technique in producing extreme ultraviolet APTs. While in the soft X-ray regime, HHG suffers from low conversion efficiency and lacking flexibility in the waveform and spectrum control. 
Here in this study, based on the waveform synthesis of the echo-enabled harmonic generation (EEHG) free-electron laser (FEL), we propose a novel method that can generate soft X-ray APTs with high temporal and spectral tunability. EEHG scheme is well-known in forming phase-locked high harmonic bunching combs down to the soft X-ray regime. And We can facilitate successive FEL lasing in several different harmonic numbers and then perform the coherent waveform synthesis to generate soft X-ray APTs. Three-dimensional simulation results indicate that reproducible generation of coherent APTs in the soft X-ray regime with peak power of 3.5~GW and micropulse duration of 160~as can be achieved by five-harmonic synthesis. And the micropulse duration together with the spectral components of the APTs can be adjusted easily according to the programmable high harmonic bunching combs in the EEHG scheme. This method has the potential to establish a robust platform for attosecond sciences in the soft X-ray regime, which can enable unprecedented studies of electron dynamics in physical and chemical reactions, biological systems and quantum materials.

\end{abstract}

\section*{Keywords}
Attosecond Waveform Synthesis, Echo-Enhance Harmonic Generation, Soft X-Ray, FEL
%%%%%% Main Text %%%%%%

\section{Introduction}

Direct observation and control of electron dynamics in their intrinsic timescales represent research frontiers across physics, chemistry, and materials science~\cite{kling2008,calegari2014}. The unprecedented fast electronic motions include photoionization delay, molecular charge migration, electron-electron correlation, electron tunneling, auger process and so on~\cite{lu2003,schultze2010,morimoto2018}. Attosecond (1~as = 10$^{-18}$~s) pulse trains (APTs) are ideal tools in the study of these phenomena, which can enable real-time tracking of fundamental quantum processes, providing unparalleled understanding on physical and chemical reactions and material properties~\cite{schafer2004,agostini2024}. High harmonic generation (HHG) technique has been successful and widely accessible in generating APTs, which is actually the enabling technique for attosecond sciences~\cite{antoine1996,paul2001}. HHG now can provide $\sim$nJ-level pulse energy per-harmonic in the extreme ultraviolet (EUV) regime, while in the soft X-ray region, HHG-based APTs face severe problems like low conversion efficiency, high medium absorption and great phase-matching difficulties~\cite{macklin1993,Gebhardt2020,Johnson2018}. These limitations are further compounded by metrological challenges in fully characterizing soft X-ray APTs~\cite{Teichmann2016}, which is an urgent need for emerging applications from attosecond near-edge X-ray absorption fine structure (NEXAFS) spectroscopy to single-shot coherent diffraction imaging~\cite{CousinHighflux2014,ReinhardSoft2025}. Extending APTs into the soft X-ray region is also essential for element-specific probing of core-level transitions, high spatiotemporal resolution imaging, and reduced radiation damage in biological samples~\cite{Fu2020,Gaumnitz2017}.

Besides HHG sources, free-electron lasers (FELs) are rapidly advancing towards the generation of APTs~\cite{maroju2020}, isolated attosecond pulses (IAPs)~\cite{PhysRevLett.119.154801,Duris.2020,Yan.2024,xiao2022,xiao2025}, and attosecond sciences~\cite{doi:10.1126/science.abj2096,doi:10.1126/science.adn6059,Driver.2024}. FELs employ high energy relativistic electron beams in advanced undulator magnets to generate ultrahigh brightness, ultrashort pulse duration, and wavelength-tunable radiation pulses from EUV to X-ray regime. Most FEL facilities are working on the basis of self‑amplified spontaneous emission (SASE) scheme~\cite{kondratenko1980,bonifacio1984}, in which the FEL longitudinal coherence is fundamentally constrained by the electron beam shot noise and the radiation slippage effects~\cite{mcneil2013}. Seeded FEL schemes such as high gain harmonic generation (HGHG)~\cite{yuGeneration1991,doi:10.1126/science.289.5481.932} and echo-enabled harmonic generation (EEHG)~\cite{stupakov2009,xiang2009} overcome the limitation and improve the longitudinal coherence by imprinting fully coherent optical laser into the electron beam. In EEHG, two laser modulators and dispersive sections are employed to form a fine-structured longitudinal phase space distribution in the electron beam, yielding a fully coherent high harmonic bunching combs with well-defined relative phases. Due to the finer bunching structures, EEHG has higher frequency-up-conversion efficiency than HGHG, which can easily reach the soft X-ray region now. Also the high harmonic bunching spectrum in EEHG-FEL is adjustable and controllable, which brings up another advantage of high spectral and temporal tunability. 

Several FEL-based methods for generating high-power APTs at short wavelength have been proposed~\cite{thompson2008,fengAttosecond2012,Xiang.2012,Dunning.2013b,maroju2021complex,maroju2020}. Among them FERMI has successfully demonstrated the APTs' generation in the EUV region~\cite{maroju2020}, by employing the waveform synthesis (WS) technique~\cite{wirth2011} into the HGHG-FEL scheme. 
In general, the WS technique is widely used in the optical community for attosecond scientific researches~\cite{Xue:22,doi:10.34133/2021/9828026}. 
And advanced study results reveal that, combining the WS technique with high gain FEL principles, APTs' generation with short pulse duration down to even zeptosecond, high photon energy beyond $10$~keV and lots of scientific applications can be made possible~\cite{Dunning.2013b,maroju2021complex,Maroju.2025}. 
The main principle of the WS technique is the coherent superimposition of multiple phase-locked optical fields. Hence it can be easily implemented in seeded FEL schemes like the HGHG and the EEHG. However, unlike in the EEHG scheme, short wavelength FEL lasing down to the soft X-ray region in the HGHG scheme is currently inaccessible. And the harmonic bunching spectrum in the HGHG scheme is largely predetermined and lacks efficient and flexible tunability.

Hence in this study, we propose a novel method to generate APTs in the soft X-ray regime with high spectral and temporal tunability by leveraging the advanced EEHG scheme and combining it with the WS technique. The unique advantage of the EEHG scheme means we can easily achieve short wavelength FEL lasing down to the soft X-ray region and even cover the "water-window" regime~\cite{rebernik2019coherent,fengCoherent2022}. And the high harmonic bunching spectrum in EEHG can be controlled intentionally and tuned easily, which helps us gain another advantage of flexible waveform and spectrum control. By strategically control and optimization of parameters like modulation strengths, chicane dispersions, and phase shifters, three-dimensional simulation results prove that, through our method, phase-locked high harmonic FEL lasing pulses in several different harmonic numbers can be obtained, coherent waveform synthesis can be performed, and soft X-ray APTs' generation can be eventually achieved. For five-harmonic synthesis, APTs in the soft X-ray region with peak power of 3.5~GW and micropulse duration of 160~as can be expected. The flexible waveform and spectrum control is also demonstrated in this study. The proposed method shows great potential in the generation and application of fully coherent soft X-ray APTs. 

\section{Methods}
%figure 1  
\begin{figure}[hb]
    \centering
    \includegraphics[width=1.0\textwidth]{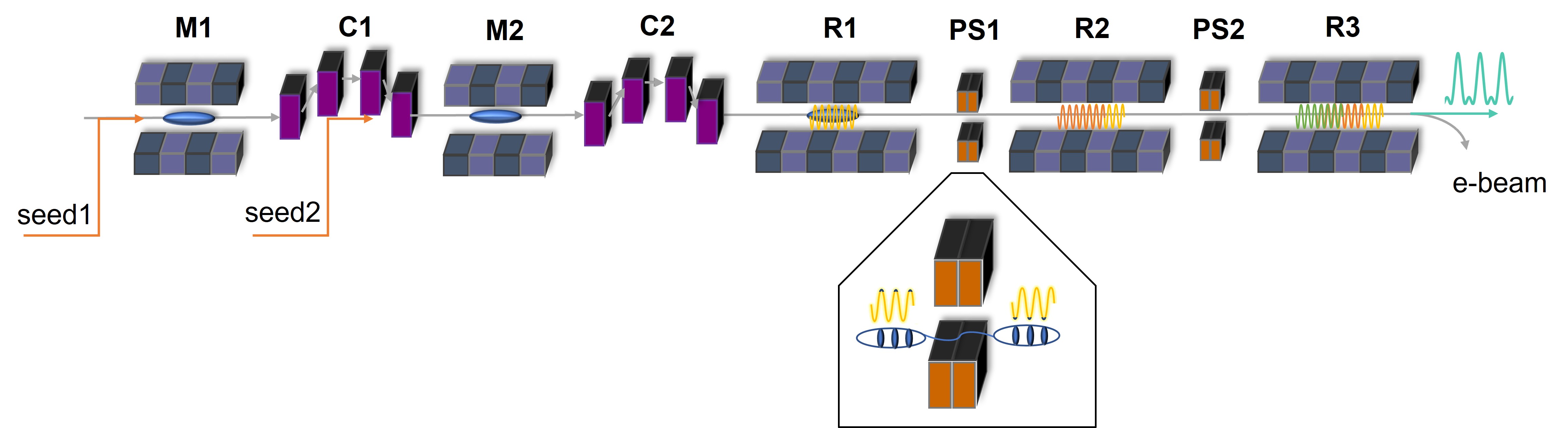}
    \caption{Schematic layout of the proposed method, mainly composing an EEHG-FEL setup and a waveform synthesis configuration.}
    \label{fig:1}
\end{figure}
The schematic layout of the proposed method is illustrated in Fig.~\ref{fig:1}, comprising three key functional sections: 
(i) two-stage modulation-dispersion sections (M1, C1, M2, C2) and two external seed lasers (Seed1, Seed2) to implement the EEHG scheme, 
(ii) multi-segment radiators (R1, R2, R3, \ldots) for FEL lasing in distinct high harmonics, and 
(iii) phase shifters (PS1, PS2,\ldots) positioned between successive radiator segments to enable precise relative phase control for the WS. 
In operation, a high-quality electron beam from the linac first traverses the EEHG scheme to acquire the high harmonic microbunching imprinted by the seed lasers. The prebunched electron beam then propagates through the radiator segments, generating coherent FEL radiation pulses at multiple harmonics with fixed phase relationships. This intrinsic phase coherence renders WS via coherent superposition feasible. Crucially, the phase shifters between radiator stages provide fine-tuning of relative phases required for stable and programmable WS, as demonstrated in prior work~\cite{shim2020}. The underlying principles and implementation details are presented in the following subsections.

\subsection{High-Harmonic Bunching Generation and Control in the EEHG scheme}

The EEHG scheme employs sequential interactions between the electron beam and external seed lasers (Seed1, Seed2) in modulators M1 and M2 to imprint energy modulations into the electron beam. The first magnetic chicane (C1), characterized by a strong longitudinal dispersion, stretches the energy modulation in M1 into finely structured energy bands in the electron beam’s longitudinal phase space. A second energy modulation is then applied in M2, superimposed on this pre-structured energy distribution. Subsequently, the weaker dispersion of chicane C2 converts the combined energy modulation into a high-fidelity density modulation. Owing to the fine granularity of the initial energy bands, this process generates pronounced microbunching at very high harmonics of the seed laser frequency.

A distinctive advantage of the EEHG scheme lies in its ability to generate very high-harmonic microbunching using only modest energy modulation amplitudes. 
The resulting harmonic bunching spectrum follows the theoretical framework established by Stupakov~\cite{stupakov2009}, Xiang~\cite{xiang2009}, and Kim~\cite{kim2017}, in which the complex bunching factor at harmonic order $h$ is given by:
\begin{equation}
     b_h = e^{i m\phi} \exp\!\left[-\frac{(h B_2 - B_1)^2}{2}\right] J_m\!\left(-h A_2 B_2\right) J_{-1}\!\left[-A_1(-B_1 + h B_2)\right],
    \label{bunching}
\end{equation}
Here $\phi$ represents the relative phase between Seed1 and Seed2, $J_m$ denotes the $m$-th order Bessel function of the first kind, 
and the harmonic number is given by $h = m\kappa - 1$ with $\kappa = \lambda_1 / \lambda_2$. Here, $\lambda_1$ and $\lambda_2$ are the wavelengths of Seed1 and Seed2, respectively, and $k_1 = 2\pi/\lambda_1$, $k_2 = 2\pi/\lambda_2$ are the corresponding wave numbers. The dimensionless modulation and dispersion parameters are defined as
\[
A_1 = \frac{\Delta E_1}{\sigma_E}, \quad A_2 = \frac{\Delta E_2}{\sigma_E}, \quad
B_1 = \frac{R_{56}^{(1)} k_1 \sigma_E}{E_0}, \quad
B_2 = \frac{R_{56}^{(2)} k_2 \sigma_E}{E_0},
\]
where $\Delta E_1$ and $\Delta E_2$ are the energy modulation amplitudes imparted in modulators M1 and M2, $R_{56}^{(1)}$ and $R_{56}^{(2)}$ are the longitudinal dispersion strengths of chicanes C1 and C2, $\sigma_E$ is the initial RMS energy spread of the electron beam, and $E_0$ is the nominal beam energy.
It follows from Eq.~\eqref{bunching} that the magnitude of the EEHG bunching factor, $|b_h|$, is primarily governed by the four dimensionless parameters $(A_1, A_2, B_1, B_2)$, enabling precise control over the harmonic content through tailored modulation and dispersion settings.

\begin{figure}[htbp]
    \centering
    \begin{subfigure}{0.32\textwidth}
        \includegraphics[width=\linewidth]{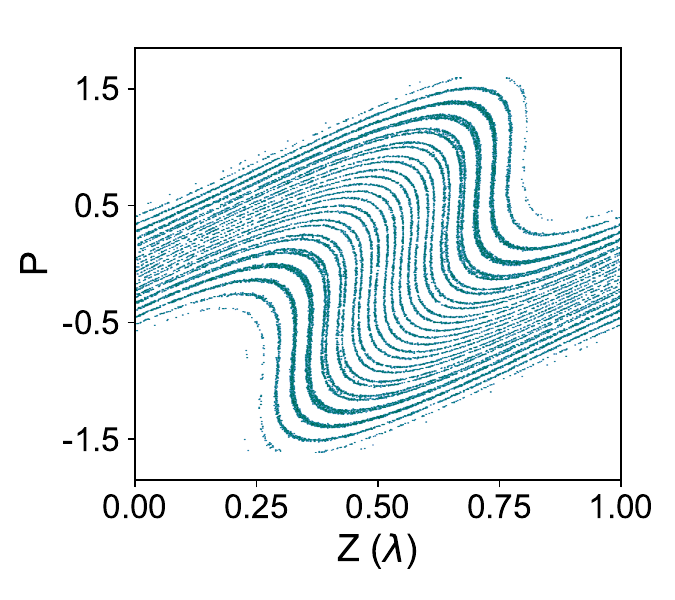}
        \caption{}
        \label{fig:eehg_dpa}
    \end{subfigure}
    % \hfill
    \begin{subfigure}{0.32\textwidth}
        \includegraphics[width=\linewidth]{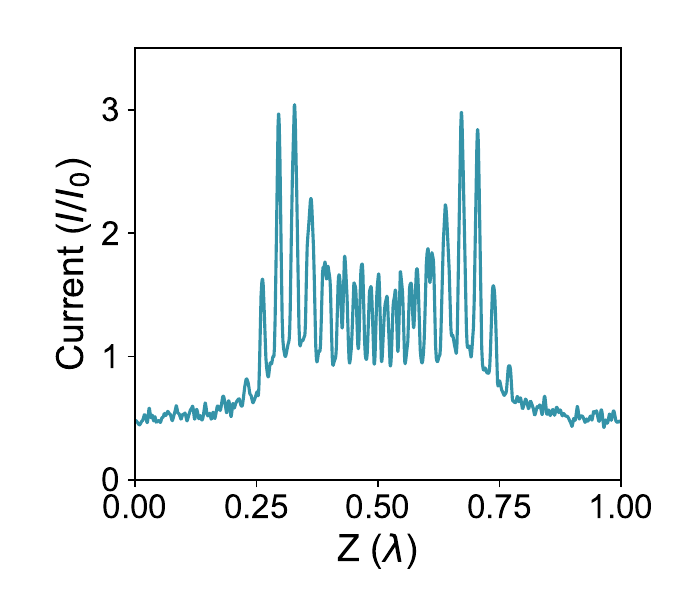}
        \caption{}
        \label{fig:eehg_cur}
    \end{subfigure}
    \begin{subfigure}{0.32\textwidth}
        \includegraphics[width=\linewidth]{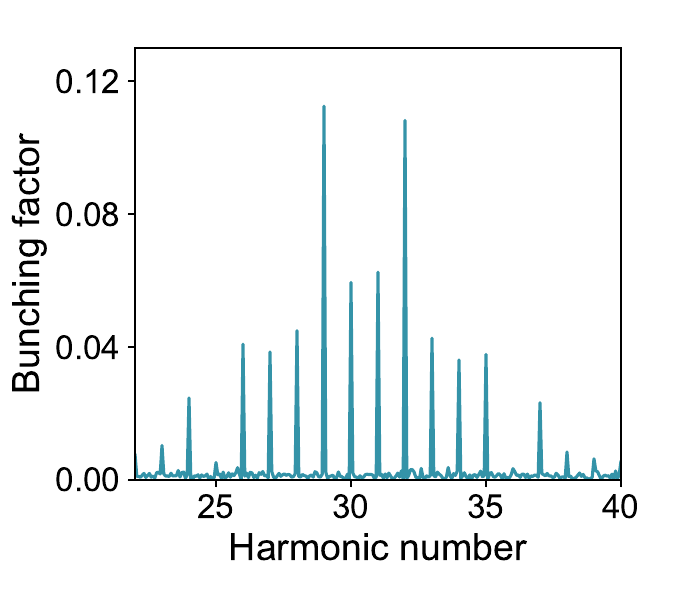}
        \caption{}
        \label{fig:eehg_har}
    \end{subfigure}
    \caption{Typical longitudinal phase space (a), current profile (b) and high harmonic bunching factor $|b_n|$ (c) for the EEHG scheme with $A_1=6.5, A_2=5.6, B_1=6.1, B_2=0.20$.}
    \label{fig:EEHGEffects}
\end{figure}
To illustrate the  characteristic effects of the EEHG scheme, we present the typical longitudinal phase-space, corresponding current profile and harmonic bunching spectrum in Fig.~\ref{fig:EEHGEffects}, simulated with modulation and dispersion parameters $A_1 = 6.5$, $A_2 = 5.6$, $B_1 = 6.1$, and $B_2 = 0.20$.  Fig.~\ref{fig:eehg_dpa} shows a longitudinal slice of the electron beam spanning one seed laser wavelength, and the corresponding current profile is shown in Fig.~\ref{fig:eehg_cur}. The fine-scale structure evident in this slice confirms the generation of very high harmonic content through the EEHG process.
The harmonic bunching spectrum is shown in Fig.~\ref{fig:eehg_har}, exhibiting a peak of nearly 11\% at the 32nd harmonic of the seed laser—achieved using only moderate modulation amplitudes ($A_1, A_2 \sim 5\text{--}6$). Notably, the EEHG bunching spectrum exhibits a distinct, non-monotonic pattern, in stark contrast to the monotonic (approximately exponential) decay observed in the HGHG scheme~\cite{yuTheory2002}. This structured spectral envelope provides an additional degree of freedom for tailoring harmonic content.

\begin{figure}[htbp]
    \centering
    \begin{subfigure}{0.45\textwidth}
        \includegraphics[width=\linewidth]{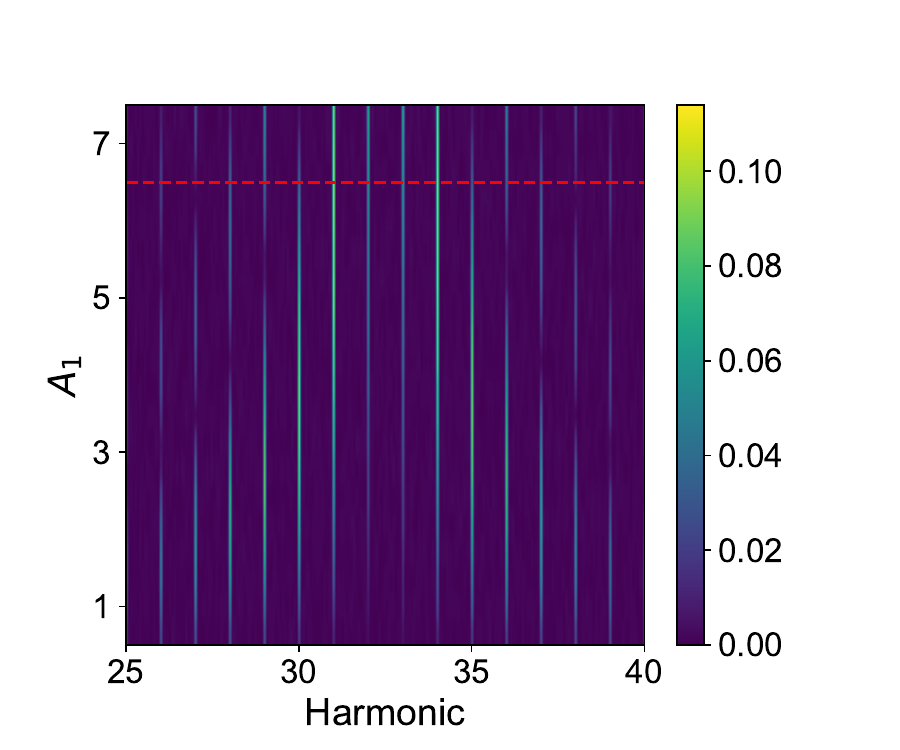}
        \caption{}
        \label{fig:bunching_evo_A1}
    \end{subfigure}
    % \hfill
    \begin{subfigure}{0.45\textwidth}
        \includegraphics[width=\linewidth]{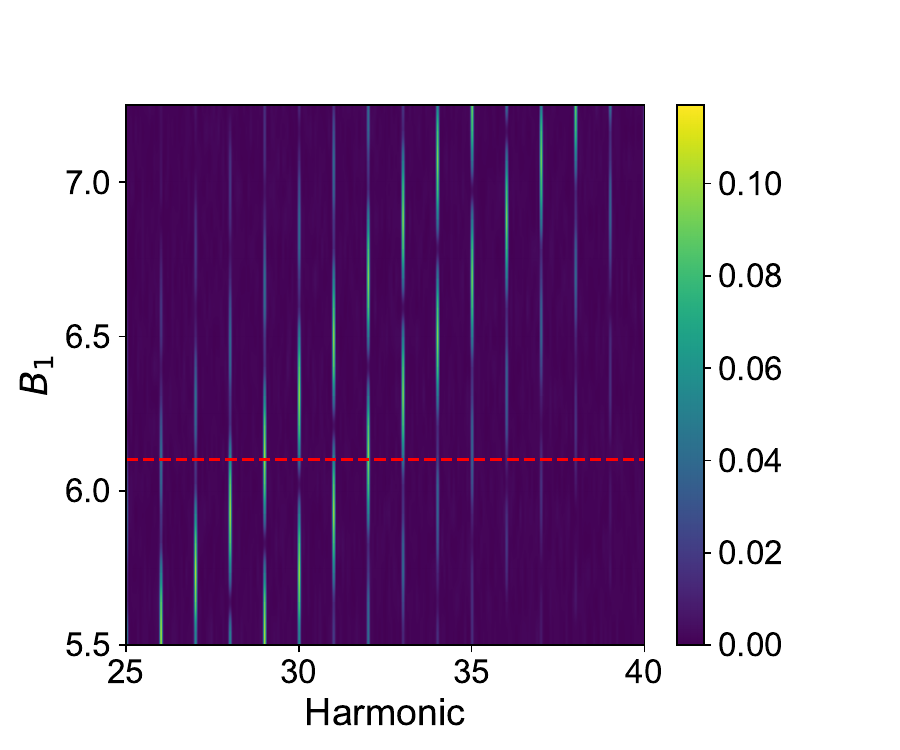}
        \caption{}
        \label{fig:bunching_evo_B1}
    \end{subfigure}
    \caption{Evolution of the harmonic bunching spectrum with the variation of EEHG parameter $A_1$ (a) and $B_1$ (b) respectively. The red dashed lines correspond to the parameter sets in Fig.~\ref{fig:EEHGEffects}.}
    \label{fig:bunching_evo}
\end{figure}
In general, the EEHG scheme enables selective enhancement of a target harmonic by optimizing the modulation and dispersion parameters. Crucially, these parameters govern distinct aspects of the harmonic bunching spectrum: the first-stage energy modulation amplitude $A_1$ primarily shapes the overall envelope and peak intensity distribution of the harmonic bunching, while the first-stage dispersion parameter $B_1$ determines the harmonic order at which maximum bunching occurs.
Fig.~\ref{fig:bunching_evo} summarizes the evolution of the harmonic bunching spectrum across harmonic orders under independent scans of $A_1$ and $B_1$. As shown in Fig.~\ref{fig:bunching_evo_A1}, decreasing the first-stage modulation amplitude $A_1$ broadens the harmonic comb and enhances bunching in the sideband harmonics, effectively increasing the spectral bandwidth of the microbunching. In contrast, increasing $A_1$ concentrates the bunching strength into fewer harmonics, yielding a narrower but more intense spectral envelope. And varying $B_1$ shifts the target harmonic with the peak bunching factor without significantly altering the overall amplitude, as illustrated in Fig.~\ref{fig:bunching_evo_B1}, which means this can enable precise tuning of the optimal harmonic number easily.
These results demonstrate that the EEHG process provides two orthogonal degrees of freedom for harmonic control: $A_1$ for spectral shaping and intensity scaling, and $B_1$ for selecting the target harmonic. This dual-parameter programmability establishes a robust foundation for flexible spectral and temporal waveform engineering in the subsequent WS processes.
 
\subsection{FEL Lasing and Harmonic Bunching Evolution in Multi-stage Radiators}
\label{sec:dispersion-enhanced}

\begin{figure}[htbp]
    \centering
    \begin{subfigure}{0.45\textwidth}
        \includegraphics[width=\linewidth]{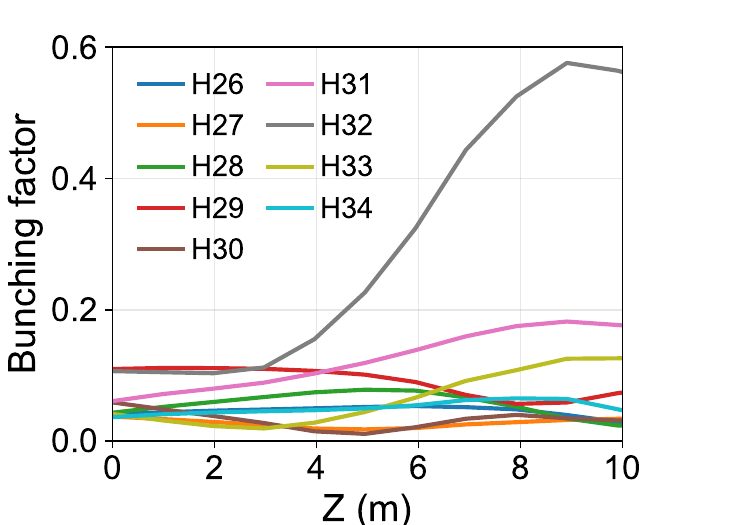}
        \caption{}
        \label{fig:bunching_evo_z}    
    \end{subfigure}
    % \hfill
    \begin{subfigure}{0.45\textwidth}
        \includegraphics[width=\linewidth]{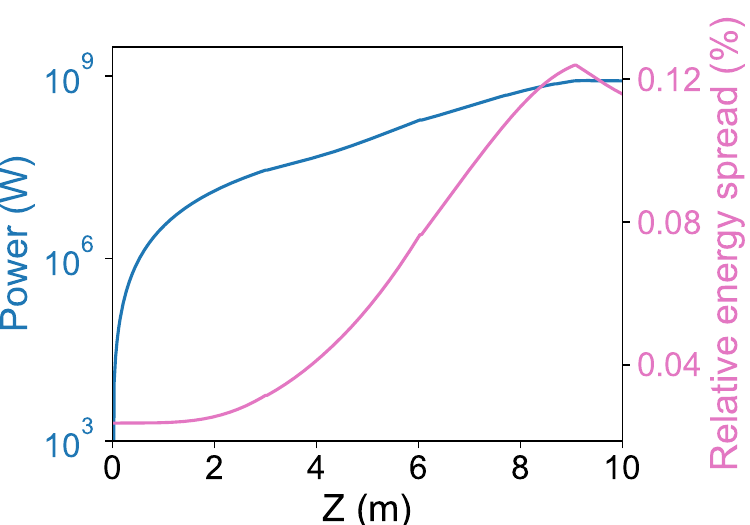}
        \caption{}
        \label{fig:powe_ens_evo}    
    \end{subfigure}
    \caption{Evolution of (a) all harmonics bunching factor and (b) the power and energy spread during the FEL lasing process. The radiator is tuned to resonate at the 32nd harmonic.}
    \label{fig:para_evo_1und}
\end{figure}
Following the generation of high-harmonic bunching combs in the EEHG stage, the electron beam propagates through a series of multi-stage radiators to undergo coherent FEL amplification at different harmonic numbers, which is prerequisite for the subsequent WS processes. Precise control over the amplification dynamics and evolution of the bunching combs in multi-stage radiators is therefore essential. Fig.~\ref{fig:para_evo_1und} illustrates a representative example of harmonic bunching evolution during the FEL lasing process in a radiator tuned to a single resonant wavelength, initialized with the same bunching spectrum as in Fig.~\ref{fig:eehg_har}. The radiator is tuned to resonance at the 32nd harmonic($H_{32}$), ensuring optimal overlap between the electron beam’s microbunching structure and the radiation field. As the beam traverses the radiator, the seeded FEL process initiates coherent energy transfer from the electrons to the radiation field at this harmonic. This interaction not only amplifies the radiation power but also reinforces the electron density modulation through the FEL ponderomotive potential—a positive feedback mechanism known as microbunching enhancement. Consequently, the bunching factor of the 32nd harmonic increases from an initial value of approximately 11\% (inherited from the EEHG stage, cf.~Fig.~\ref{fig:eehg_har}) to over 58\% at saturation, while the FEL power grows to the gigawatt level and the relative energy spread rises from 0.02\% to 0.12\%. This coherent exponential gain process significantly enables high power FEL lasing and strengthens the electron beam microbunching.  

In addition, Fig.~\ref{fig:bunching_evo_z} reveals that the initially strong bunching factor of the 30th harmonic ($H_{30}$) gradually decreasing and the bunching factor of 31st harmonic ($H_{31}$) gradually increasing to nearly 15\% at $z = 8$~m. The reason is that, while the initial harmonic bunching spectrum is tailored by the modulation and dispersion parameters ($A_1$, $A_2$, $B_1$, $B_2$), the longitudinal dispersion $R_{56}$ introduced in the radiator further modulates the bunching evolution. Specifically, this radiator dispersion preferentially amplifies lower-order harmonics near the resonant frequency (here, $H_{32}$) and facilitates smooth gain transfer among neighboring harmonics. This dispersion-engineered bunching dynamics enables the sustained amplification of multiple high harmonics along the radiator line. Consequently, in a multi-stage radiator configuration, successive segments can be tuned to lase at different harmonic numbers—such as $H_{32}$, $H_{31}$, and $H_{30}$—producing a set of phase-locked radiation pulses. These pulses will serve as the essential building blocks for WS and the subsequent generation of APTs. 

It is important to note that the significant growth in FEL radiation power will inevitably increase the electron beam’s energy spread. This enlarged energy spread will hence degrade the microbunching fidelity and suppress further FEL amplification in downstream radiator sections. Consequently, for each harmonic lasing stage, both the saturation power and the radiator length must be carefully optimized to balance high output power against beam quality preservation. Furthermore, the longitudinal dispersion $R_{56}$ that can be practically introduced in the radiator segments is inherently limited by the overall saturation length, and excessive dispersion would distort the electron beam’s dynamics. This constraint restricts the spectral window over which neighboring harmonics can be simultaneously amplified through dispersion engineering, thereby defining an upper bound on the harmonic comb bandwidth achievable in a single radiator stage.

\begin{figure}[htbp]
    \centering
    \includegraphics[width=0.9\linewidth]{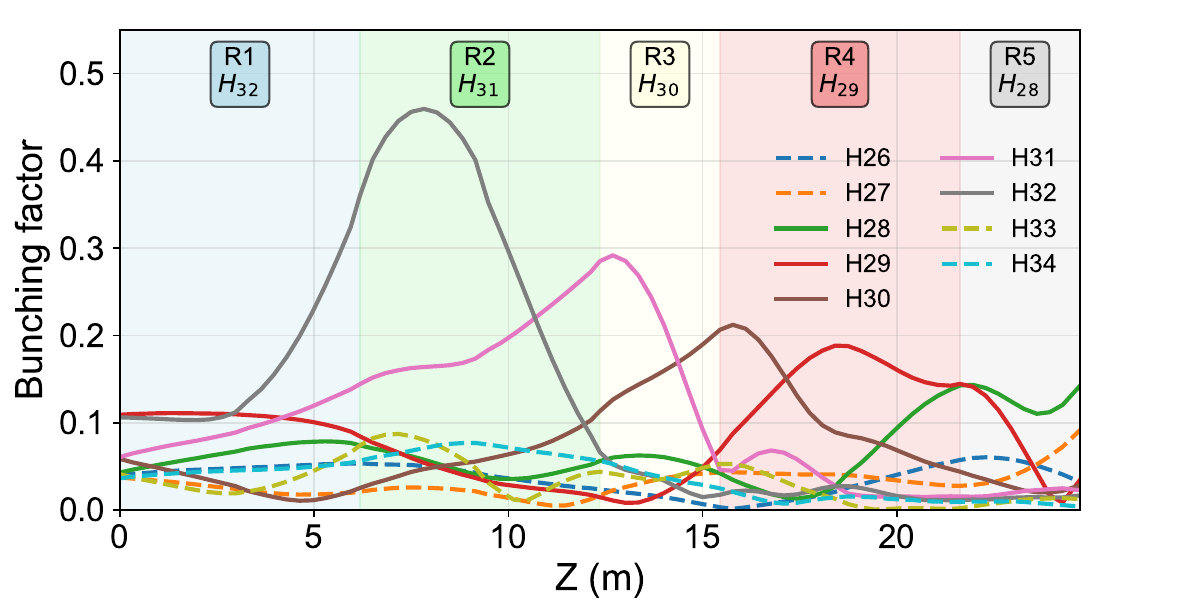}
    \caption{ Dispersion-enhanced high harmonic bunching evolution along the five-segment radiator system for five-harmonic synthesis.}
    \label{fig:bunching_evo_z_Sn5}
\end{figure}
Based on the principles and analysis outlined above,  a five-segment radiator system—tuned sequentially to $H_{32}$ through $H_{28}$ with tailored segment lengths—can be implemented. The evolution of harmonic bunching across these radiator sections is shown in Fig.~\ref{fig:bunching_evo_z_Sn5}. As previously discussed, the bunching factor of lower-order harmonics near the resonant harmonic is enhanced during FEL amplification, enabling their subsequent lasing in downstream radiator stages. 
For instance, radiator section R1 is tuned to ($H_{32}$). During FEL amplification in R1, not only does the bunching factor of $H_{32}$ grow significantly, but that of the adjacent $H_{31}$ also increases substantially. This enhanced $H_{31}$ microbunching then serves as the seed for FEL lasing in the next radiator section, R2. This cascaded amplification process can be repeated across multiple stages, provided that the bunching structure remains intact and the electron beam retains sufficient coherence.
However, the process eventually terminates when either the EEHG-induced microbunching structure is disrupted, causing the bunching factor of the next target harmonic to become negligible, or the energy spread of the electron beam is significantly increased by prior FEL interactions, thereby suppressing further amplification.
These results demonstrate that judicious dispersion management, combined with multi-stage FEL amplification, enables flexible, configuration-dependent shaping of the harmonic combs. Specifically, the synergistic integration of the EEHG parameters $(A_1, A_2, B_1, B_2)$, longitudinal dispersion $R_{56}$, and tailored multi-segment radiators collectively facilitate dynamic control over high-harmonic bunching evolution. This capability constitutes the core methodological innovation underlying the programmable soft X-ray WS presented in the following sections.

\subsection{Coherent Multicolor WS for APTs' Generation}
\label{sec:synthesis}
%●	简述利用多个相干谐波分量干涉形成阿秒脉冲串的基本概念。
%●	强调合成脉冲特性（脉宽、强度、包络）由参与合成的谐波分量的数量、振幅和相对相位决定。

With a programmable harmonic comb and multi-harmonic FEL lasing established, we can then perform the WS technique to generate APTs through the coherent superposition of the radiation fields. The temporal profile of the synthesized pulses is governed by four controllable parameters: the number of harmonics (\(N\)), their spectral amplitudes (\(M_k\)), angular frequencies (\(\omega_k\)), and phases (\(\phi_k\)). Precise phase matching among these harmonics is achieved via phase shifters integrated between successive radiator segments. This is essential for stable WS. The phase matching is achieved by tuning the deflection parameter $K$ of the phase shifters. The resulting phase shift $\Delta\phi$ is given by
\begin{equation}
    \Delta\phi = \frac{2\pi}{\lambda} \frac{L_{\mathrm{ps}}}{2\gamma^2}\left(1 + \frac{K^2}{2}\right),
\end{equation}
where $\lambda$ is the resonant wavelength of the target harmonic, $L_{\mathrm{ps}}$ denotes the effective length of the phase shifter, and $\gamma$ is the relativistic Lorentz factor. 
The phase control precision of $\Delta\phi$ can reach below 0.1~rad, which can fully satisfy the requirements of the proposed method.

To elucidate the underlying generation mechanism, we define the instantaneous radiation intensity as \(I_{\text{inst}}(t)\). According to Poynting’s theorem, this intensity can be expressed as
\begin{equation}
    I_{\text{inst}}(t) = \epsilon_0 c \, E_{\text{total}}^2(t),
\end{equation}
where the total electric field comprises \(N\) harmonic components:
\begin{equation}
    E_{\mathrm{total}}(t) = \sum_{k=1}^{N} M_{k} \, e^{i\left(\omega_{k} t + \phi_{k}\right)}.
\end{equation}
For closely spaced harmonics (\(\Delta\omega \ll \bar{\omega}\)), the rapidly oscillating carrier terms average out over observation timescales \(\tau \gg 2\pi/\bar{\omega}\), yielding the slowly varying envelope intensity:
\begin{equation}
    I_{\mathrm{env}}(t) = \frac{\epsilon_0 c}{2} \left[ \sum_{k=1}^{N} M_k^2 + 2 \sum_{i<j} M_i M_j \cos\!\left(\Omega_{ij} t + \Phi_{ij}\right) \right],
    \label{I_env}
\end{equation}
where \(\Omega_{ij} = \omega_i - \omega_j\) and \(\Phi_{ij} = \phi_i - \phi_j\) denote the frequency and phase differences between harmonics \(i\) and \(j\), respectively.

The maximum intensity is attained when all interference terms constructively interfere, i.e., when \(\cos(\Omega_{ij} t + \Phi_{ij}) = 1\) for all \(i < j\):
\begin{equation}
    I_{\mathrm{max}} = \frac{\epsilon_0 c}{2} \left( \sum_{k=1}^{N} M_k \right)^2.
    \label{I_max}
\end{equation}
This condition requires the relative phases to satisfy a linear relationship with the frequency differences, which means equivalently a constant group delay (i.e., zero group-delay dispersion) across the harmonic comb:
\begin{equation}
    \frac{\Phi_{ij}}{\Omega_{ij}} = \tau_0 \quad (\tau_0 \in \mathbb{R}),
    \label{phi}
\end{equation}
which will ensure temporal synchronization of all harmonic components. Under this phase-matching condition, the synthesized waveform becomes strictly periodic, with pulse period determined by the greatest common divisor of all pairwise frequency differences:
\begin{equation}
    T = \frac{2\pi}{\gcd\!\left(\{\Omega_{ij}\}\right)}.
    \label{T}
\end{equation}

To quantify the degree of coherent enhancement achieved through WS, we define the peak intensity enhancement factor as
\begin{equation}
    \alpha = \frac{I_{\mathrm{max}}}{\sum_{k=1}^{N} I_k} = \frac{\left( \sum_{k=1}^{N} M_k \right)^2}{\sum_{k=1}^{N} M_k^2},
    \label{alpha}
\end{equation}
where \(I_k = \frac{\epsilon_0 c}{2} M_k^2\) is the intensity of the \(k\)-th harmonic. The enhancement factor reaches its theoretical maximum \(\alpha_{\mathrm{max}} = N\) when all amplitudes are equal (\(M_k = M\)).
Under ideal phase matching, the minimum intensity can vanish if the largest amplitude does not dominate the sum of the others, i.e., if \(M_{\mathrm{max}} \leq \sum_{k \neq \mathrm{max}} M_k\). Otherwise, the intensity is bounded below by
\begin{equation}
    I_{\mathrm{min}} \geq \frac{\epsilon_0 c}{2} \left( 2M_{\mathrm{max}} - \sum_{k=1}^{N} M_k \right)^2.
    \label{I_min}
\end{equation}

For dual-harmonic synthesis (\(N=2\)), the relative phase \(\Phi_{12} = \phi_1 - \phi_2\) can be freely adjusted, and the optimal signal-to-noise ratio (SNR)—defined as the peak-to-sidelobe intensity ratio—is achieved when \(M_1 = M_2\), yielding a vanishing minimum intensity and maximal \(\alpha = 2\).
While in contrast, multi-harmonic synthesis (\(N \geq 3\)) requires strict adherence to the linear phase-matching condition in Eq.~\eqref{phi}. While the peak intensity enhancement factor \(\alpha\) is maximized for equal amplitudes (\(M_k = M\)), the temporal contrast (SNR) can be further optimized through tailored amplitude shaping. Moreover, equally spaced harmonics are essential to ensure a periodic pulse train with a well-defined repetition rate and high temporal contrast.

\section{Results}
\subsection{Soft X-ray APTs' Generation}
\label{sec:five}

\begin{table}[htbp]
    \caption{Simulation parameters based on the SXFEL facility.}    
    \centering
    \begin{tabular}{lcc}
        \hline
        Parameter & Value & Unit \\  
        \hline
        \textbf{Electron Beam}  \\ 
        Energy & 1.3 & GeV \\ 
        RMS energy spread & 60 & keV \\
        Normalized emittance & 1 & mm·mrad \\
        Peak current & 800 & A \\
        \textbf{Seed Lasers}   \\
        Wavelength & 266 & nm \\
        Pulse duration (Seed1) & 100 & fs \\
        Pulse duration (Seed2) & 100 & fs \\
        \textbf{Undulator System} \\  
        Modulator period (M1, M2) & 8 & cm \\ 
        Modulator length (M1, M2) & 1.28 & m \\
        Radiator period & 3 & cm \\
        Radiator segment length & 3 & m \\  
        K (R1) & 1.61 & -- \\
        K (R2) & 1.64 & -- \\
        K (R3) & 1.68 & -- \\
        K (R4) & 1.72 & -- \\
        K (R5) & 1.76 & -- \\
        \hline
    \end{tabular}
    \label{tab:1}
\end{table}
To evaluate the proposed method, we performed three-dimensional numerical simulations based on realistic machine parameters from the Shanghai Soft X-ray Free-Electron Laser (SXFEL) facility~\cite{liu2021}. The FEL amplification process was modeled using the time-dependent FEL code \textsc{Genesis 1.3}~\cite{reiche1999}. The electron beam parameters were set to representative operational values: a beam energy of 1.3~GeV, an RMS energy spread of 60~keV, a normalized emittance of 1~mm·mrad, and a peak current of 800~A. Two external seed lasers at 266~nm were employed, Seed1 and Seed2, both with a pulse duration of 100~fs, were derived from the same laser source. The undulator system was optimized for soft X-ray harmonic generation, featuring a period of 3~cm and alternating radiator segments of 3~m length. A complete list of simulation parameters is provided in Table~\ref{tab:1}. 

\begin{figure}[htbp]
    \centering
    \begin{subfigure}{0.32\textwidth}
        \includegraphics[width=\linewidth]{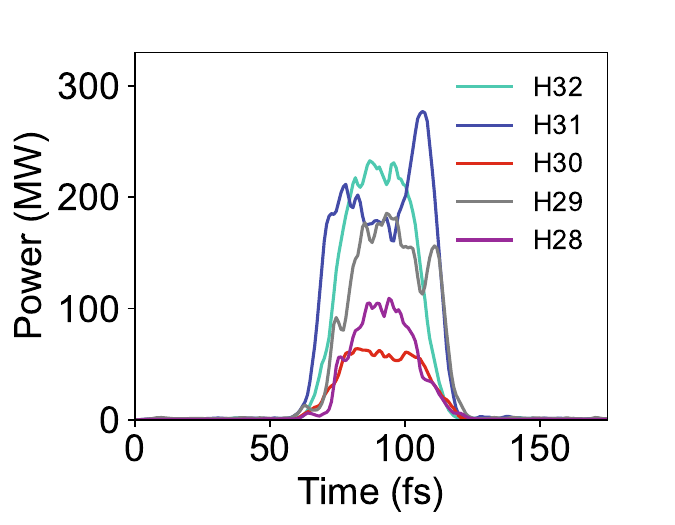}
        \caption{}
        \label{fig:power_radn5}
    \end{subfigure}
    \hfill
    \begin{subfigure}{0.32\textwidth}
        \includegraphics[width=\linewidth]{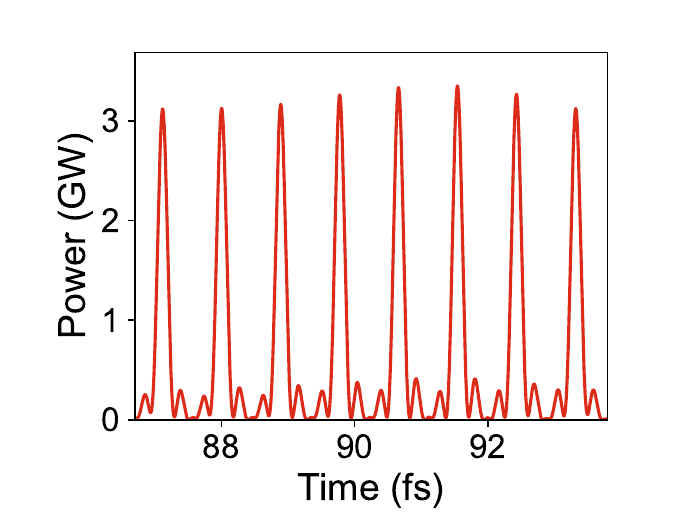}   
        \caption{}
        \label{fig:power_syn_rad5}
    \end{subfigure}
    \hfill
    \begin{subfigure}{0.32\textwidth}
        \includegraphics[width=\linewidth]{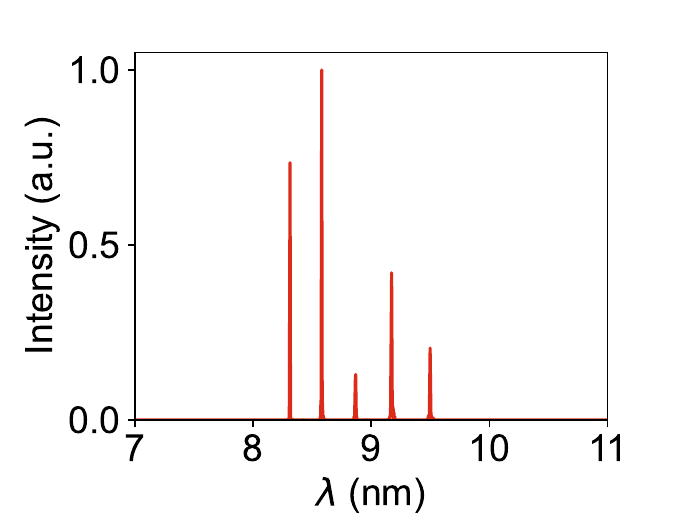}
        \caption{}
        \label{fig:spec_syn_rad5}
    \end{subfigure}
    \caption{Five-harmonic WS results for consecutive harmonics from $H_{32}$ to $H_{28}$: (a)~radiation power profile of individual harmonic number, (b)~temporal profile, and (c)~phase-locked harmonic components of the synthesized APTs.}
    \label{fig:all_syn_rad5}
\end{figure}

We simulated a five-harmonic WS scheme spanning consecutive harmonics from the $H_{32}$ to the $H_{28}$. The EEHG stage was configured with the same parameters as in Fig.~\ref{fig:eehg_har}, yielding a broad and structured bunching comb with significant amplitudes across the target harmonic range. This initial harmonic bunching spectrum serves as the seed for staged amplification in a five-segment radiator system, whose layout is illustrated in Fig.~\ref{fig:bunching_evo_z_Sn5}. As the pre-bunched electron beam propagates through successive radiator sections—each tuned to resonate at a specific harmonic (R1: $H_{32}$, R2: $H_{31}$, R3: $H_{30}$, R4: $H_{29}$, R5: $H_{28}$)—the interplay between FEL gain and radiator-induced longitudinal dispersion ($R_{56}$) enables efficient energy extraction into multiple harmonic radiation fields, which means we can get FEL lasing pulses in several different harmonics. Fig.~\ref{fig:power_radn5} shows the results of the radiation power profiles at the exit of each radiator section, confirming that all five harmonics FEL radiation pulses reach tens to hundreds of megawatts peak power output. This successful multi-harmonic FEL lasing is a direct manifestation of the dispersion-managed bunching evolution described in Sec.~\ref{sec:dispersion-enhanced}, where $R_{56}$ in each radiator selectively enhances lower-order harmonics near resonance, thereby facilitating smooth gain handover between successive stages.

The phase coherence imprinted during the EEHG process is preserved throughout the amplification chain, as evidenced by the narrow spectral linewidths of the harmonic fields in Fig.~\ref{fig:spec_syn_rad5}. This coherence will enable high-fidelity coherent superposition via the WS technique. And all the spectral lines are within the soft X-ray regime. The four phase shifters between the five radiator segments were adjusted to satisfy the linear phase-matching condition in Eq.~\eqref{phi} with $\Delta\Phi_{ij} = \phi_{12} = \phi_{23} = \phi_{34} = \phi_{45}= 0$ (the parameter will become the default in simulations unless stated otherwise), yielding a synthesize APTs shown in Fig.~\ref{fig:power_syn_rad5} exhibit a compressed micro-pulse duration of 160~as, a repetition rate of $2.25 \times 10^{15}$~Hz, and a peak power exceeding 3.5~GW. The temporal contrast being nearly 10, defined as the ratio of the main pulse peak intensity to that of the nearest sidelobe, confirms that the synthesized waveform is not only intense but also highly structured, which will fulfill the stringent requirements for phase-sensitive attosecond spectroscopy~\cite{paul2001}.
The results demonstrate that by synergistically combining programmable EEHG bunching, dispersion-engineered radiator staging, and precise phase control, our method can achieve simultaneous manipulation and FEL lasing of several different harmonics, which is a critical step toward fully customizable soft X-ray APTs.

\subsection{Waveform and Spectrum Control}
\begin{figure}[htbp]
    \centering
    \begin{subfigure}{0.45\textwidth}
        \includegraphics[width=\linewidth]{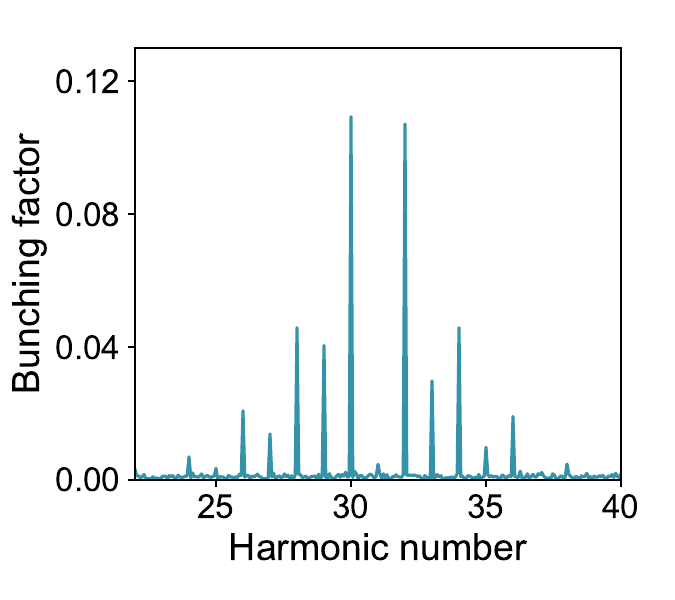}
        \caption{}
        \label{fig:bunching_rad3}
    \end{subfigure}
    \begin{subfigure}{0.45\textwidth}
        \includegraphics[width=\linewidth]{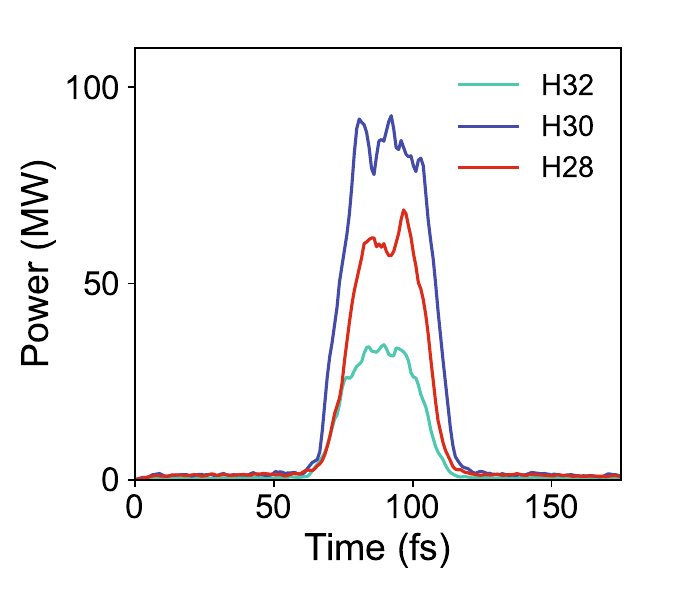}
        \caption{}
        \label{fig:radn3}
    \end{subfigure}
    \caption{High harmonic bunching factor (a) and radiation power of individual harmonics (b) for non-consecutive harmonic synthesis with $H_{32}$, $H_{30}$, and $H_{28}$. The EEHG parameters are tuned to be $A_1=6.0, A_2=4.0, B_1=8.6, B_2=0.28$ now.}
    \label{fig:all_bunching_rad3}
\end{figure}

The above results demonstrate successful waveform synthesis and soft X-ray APTs' generation from a consecutive five harmonic combs. To further highlight the flexibility of waveform and spectrum control in our method, we consider a non-consecutive harmonic comb by re-optimizing the EEHG parameters to enhance bunching at selected target harmonics. 
Specifically, we demonstrate triple-harmonic WS by coherently combining the $H_{32}$, $H_{30}$, and $H_{28}$ to further validate the waveform tunability in our method. The modulation amplitudes were set to $A_1 = 6.0$ and $A_2 = 4.0$, with normalized dispersion strengths $B_1 = 8.6$ and $B_2 = 0.28$, respectively. This configuration yields pronounced initial microbunching at $H_{32}$ and $H_{30}$, with bunching factors exceeding 0.11, as shown in Fig.~\ref{fig:bunching_rad3}.

\begin{figure}[t]
    \centering
    \begin{subfigure}{0.45\textwidth}
        \includegraphics[width=\linewidth]{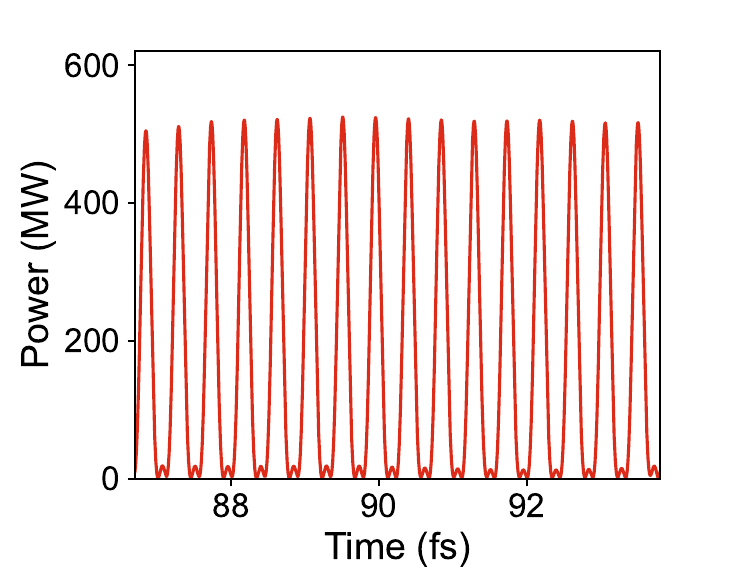}
        \caption{}
        \label{fig:power_syn_rad3}
    \end{subfigure}
    % \hfill
    \begin{subfigure}{0.45\textwidth}
        \includegraphics[width=\linewidth]{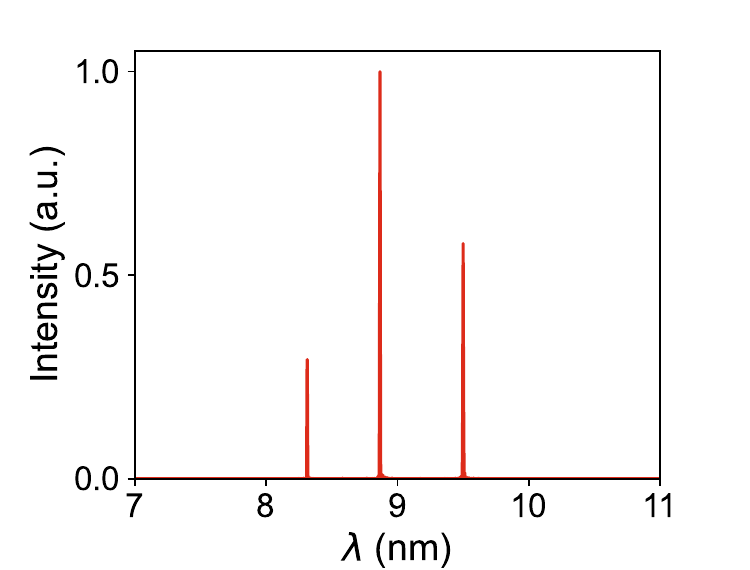}
        \caption{}
        \label{fig:spec_syn_rad3}
    \end{subfigure}
    \vspace{1mm}
    \begin{subfigure}{0.45\textwidth}
        \includegraphics[width=\linewidth]{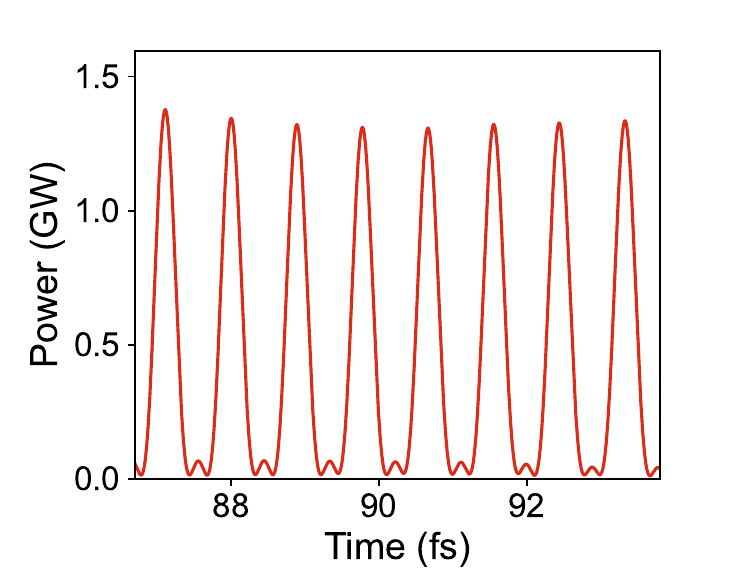}
        \caption{}  
        \label{fig:power_syn_rad3_h1}
    \end{subfigure}
    \begin{subfigure}{0.45\textwidth}
        \includegraphics[width=\linewidth]{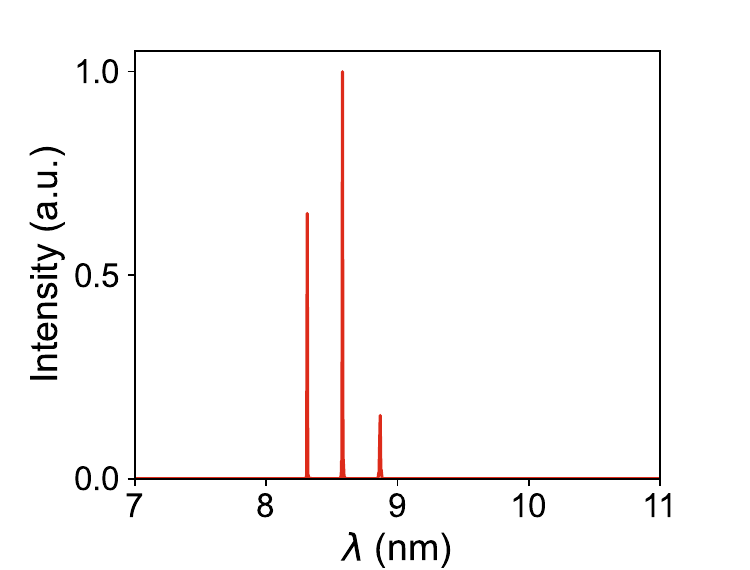}
        \caption{}  
        \label{fig:spec_syn_rad3_h1}
    \end{subfigure}
    \caption{ Power and spectrum profile of the synthesized APTs for triple-harmonic synthesis with non-consecutive $H_{32}$, $H_{30}$, $H_{28}$  (a,b) and consecutive $H_{32}$, $H_{31}$, $H_{30}$ (c,d) harmonic numbers.}
    \label{fig:all_syn_rad3}
\end{figure}

In the first radiator segment (a single-undulator, 3~m), the 32nd harmonic reached a peak power of 35~MW. The 30th harmonic reached a peak power of 85~MW in the second radiator segment (a dual-undulator, 6~m). The third radiator segment also employed a 6~m dual-undulator configuration to enhance radiation efficiency, achieving a peak power of 65~MW at the 28th harmonic. The two phase shifters between the three radiator segments were adjusted to satisfy the linear phase-matching condition in Eq.~\eqref{phi}, yielding a synthesized APTs with a peak power of 525~MW, a micropulse duration of 190~as, and a temporal contrast of 25. The corresponding temporal profile and phase-locked spectrum are shown in Fig.~\ref{fig:power_syn_rad3} and~\ref{fig:spec_syn_rad3} respectively. 
For comparison purpose, we also illustrate the synthesized results of three consecutive harmonics ($H_{32}$–$H_{30}$) extracted from the first three radiator stages in Fig.~\ref{fig:all_syn_rad5}. The temporal and spectral results are shown in Figs.~\ref{fig:power_syn_rad3_h1} and~\ref{fig:spec_syn_rad3_h1}, respectively. 
One can find out that the micropulse duration here is clearly increased, which is about 335~as comparing to the 190~as in Fig.~\ref{fig:power_syn_rad3}. The results presented here means we can adjust the waveform of the APTs by synthesis of non-consecutive harmonic combs in our method.
% One can find out that the micropulse duration here is clearly increased, which is about 335~as comparing to the 190~as in Fig.~\ref{fig:power_syn_rad3}. 
% The results presented here means we can adjust the waveform of the APTs by synthesis of non-consecutive harmonic combs in our method.    

\begin{figure}[t]
    \centering
    \begin{subfigure}{0.45\textwidth}
        \includegraphics[width=\linewidth]{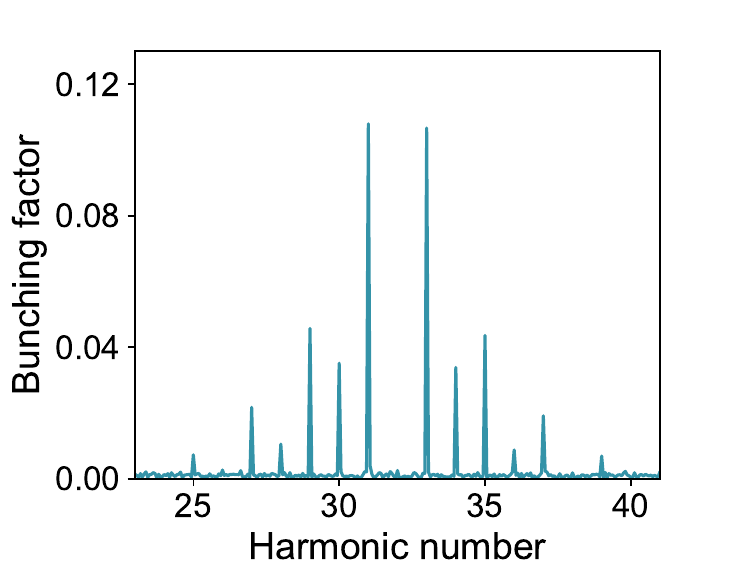}
        \caption{}
        \label{fig:bunching_rad3_offset}
    \end{subfigure}
    \begin{subfigure}{0.45\textwidth}
        \includegraphics[width=\linewidth]{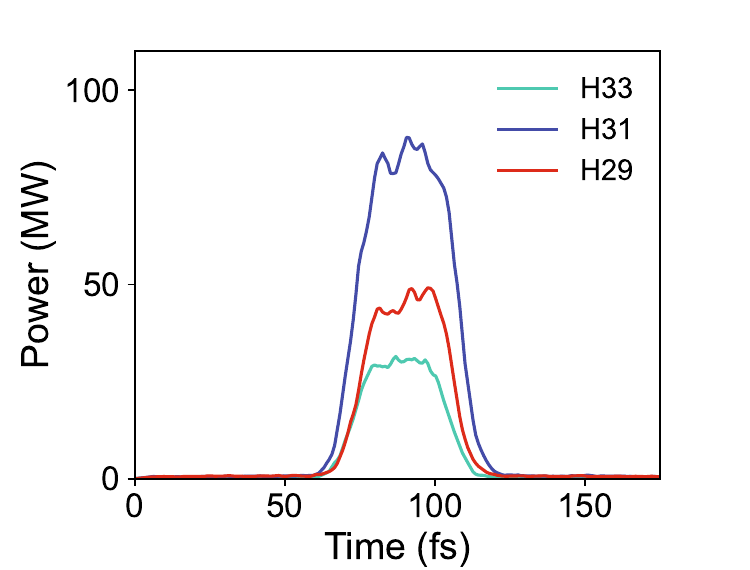}
        \caption{}
        \label{fig:power_rad3_offset}
    \end{subfigure}
        \vspace{1mm}
    \begin{subfigure}{0.45\textwidth}
        \includegraphics[width=\linewidth]{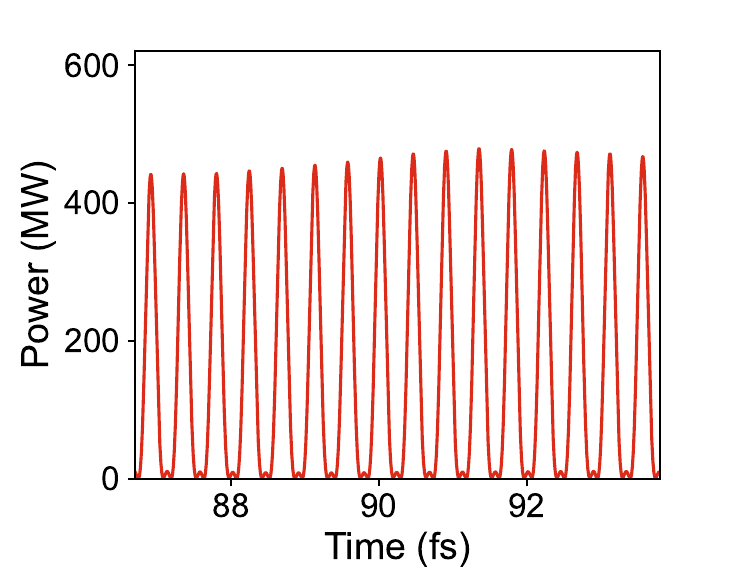}
        \caption{}
        \label{fig:power_syn_rad3_offset}
    \end{subfigure}
    % \hfill
    \begin{subfigure}{0.45\textwidth}
        \includegraphics[width=\linewidth]{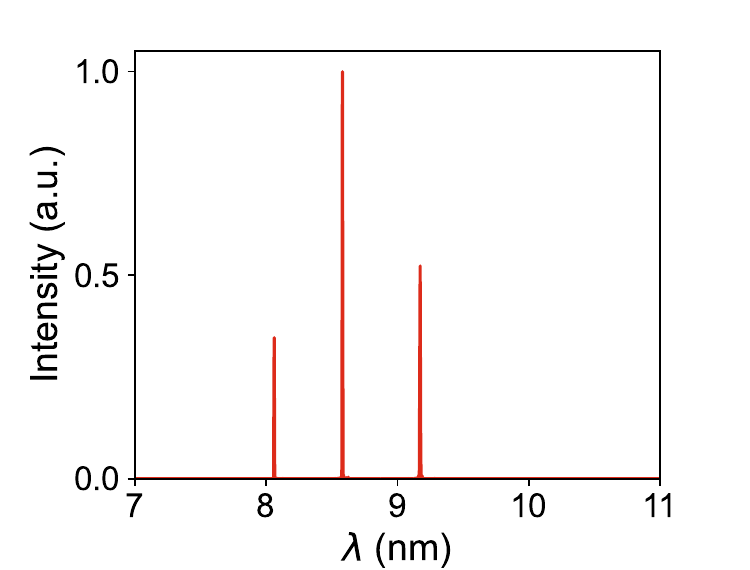}
        \caption{}
        \label{fig:spec_syn_rad3_offset}
    \end{subfigure}
    \caption{High harmonic bunching factor~(a), radiation power of each harmonic~(b), and the synthesized power (c) and spectrum (d) profile when shifting the spectrum one-harmonic-order larger by varying $B_1$ from 8.6 to 8.9 in the EEHG scheme.}
    \label{fig:all_rad3_offset}
\end{figure}

To further validate the spectral tunability, we adjusted the EEHG dispersion parameter $B_1$ from 8.6 to 8.9 while retaining the non-consecutive harmonic layout. We know from Fig.~\ref{fig:bunching_evo_B1} that, variation of $B_1$ will shift the high harmonic bunching spectrum in the EEHG scheme. And the modified initial bunching spectrum now is shown in Fig.~\ref{fig:bunching_rad3_offset}, in which we can see the optimal bunching harmonics are $H_{33}$ and $H_{31}$, which is shifted one-harmonic-order larger from $H_{32}$ and $H_{30}$ in Fig.~\ref{fig:bunching_rad3}.
Keeping the radiator architecture unchanged, we then retuned the resonance conditions of the three radiator segments to $H_{33}$, $H_{31}$, and $H_{29}$, respectively. And the power distribution of high harmonics FEL lasing at the exit of each radiator segment is presented in Fig.~\ref{fig:power_rad3_offset}. A comparison with the case in Fig.~\ref{fig:radn3} reveals that the output powers of all three harmonics remain the same levels, confirming successful amplification across the shifted harmonic combs. And after adjusting the phase shifters here, we can also achieve coherent WS of the $H_{33}$, $H_{31}$, and $H_{29}$ harmonics. The resulting temporal and spectral characteristics are shown in Fig.~\ref{fig:power_syn_rad3_offset},~\ref{fig:spec_syn_rad3_offset}. 
Comparison with the baseline synthesis results in Fig.~\ref{fig:power_syn_rad3},\ref{fig:spec_syn_rad3}, one can find out that the synthesized waveform retains a nearly identical pulse structure and micropulse duration, while the entire spectrum is shifted to shorter wavelengths accordingly. These results prove that the proposed scheme offers excellent flexibility in waveform and spectrum control over the soft X-ray APTs' generation.

\section{Discussion and Conclusion}
\begin{figure}[htbp]
    \centering
    \begin{subfigure}{0.45\textwidth}
        \includegraphics[width=\linewidth]{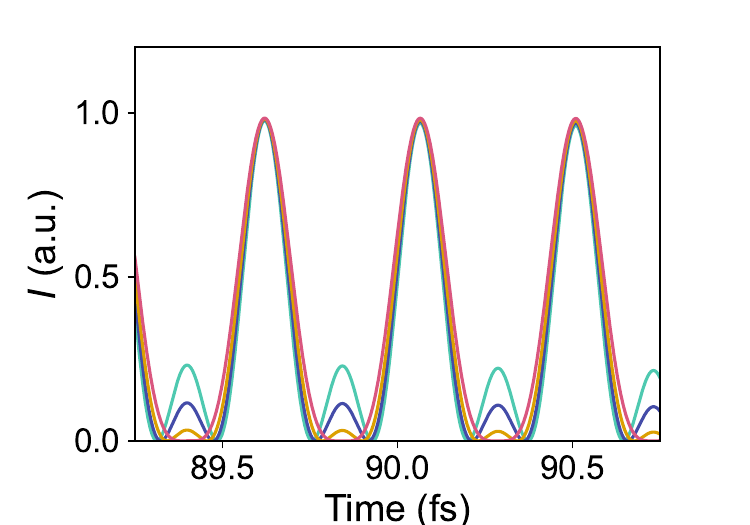}
        \caption{}
        \label{fig:power_syn_rad3_A_phi0}
    \end{subfigure}
    \hspace{1mm}
    \begin{subfigure}{0.45\textwidth}
        \includegraphics[width=\linewidth]{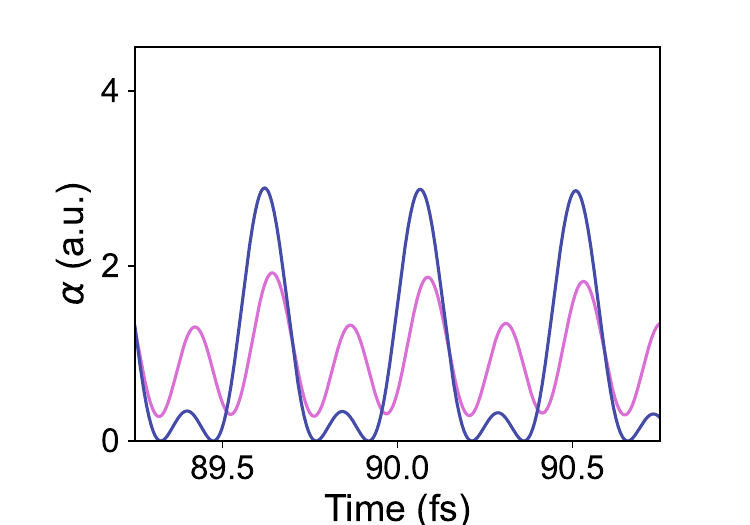}
        \caption{}
        \label{fig:power_syn_rad3_A_phix}
    \end{subfigure}
    \caption{The influences of the different relative intensity ratios (a) and the phase-matching conditions (b) on the synthesized APTs. In panel (a) the relative intensities are $1:0.5:1$ (light blue curve), $1:1:1$ (blue curve), $1:2:1$ (yellow curve) and $1:4:1$ (red curve) respectively. In panel (b) the blue curve is under the phase-matching condition while the magenta curve is not.}
    \label{fig:power_syn_rad3_diff}
\end{figure}

%Include a Discussion that summarizes (but does not merely repeat) your conclusions and elaborates on their implications. There should be a paragraph outlining the limitations of your results and interpretation, as well as a discussion of the steps that need to be taken for the findings to be applied. Please avoid claims of priority. 
In general, we demonstrate a novel method that can generate soft X-ray APTs by combining the WS technique with the EEHG-FEL scheme. The waveform and spectrum of the APTs can also be controlled and adjusted easily taking advantage of the programmable high harmonic bunching combs in the EEHG scheme.
We want to emphasize that the fully coherent properties of each harmonic component in the synthesized APTs are preserved perfectly by the intrinsic seeded EEHG-FEL principle. However, the APTs' temporal contrast is apparently sensitive to the relative intensity distributions among the harmonic components. To evaluate the influences, We present a further analysis results in Fig.~\ref{fig:power_syn_rad3_A_phi0} based on the non-consecutive harmonic synthesis ($H_{32}:H_{30}:H_{28}$). The relative intensities are set to be $1:0.5:1$ (light blue curve), $1:1:1$ (blue curve), $1:2:1$ (yellow curve) and $1:4:1$ (red curve) respectively. One can find out that, increasing the relative weight of the central harmonic ($H_{30}$) will enhance the temporal contrast. When the intensity ratio is tuned to $1:4:1$, the sidelobes are nearly suppressed to zero, yielding the highest signal-to-noise (SNR) ratio.
However, this will also result in the pulse broadening in the main peak. And if the central harmonic intensity is too high and weight exceeds this condition, the former sidelobe positions evolve into a quasi-continuous power pedestal, causing adjacent pulses in the train to merge and lose temporal isolation.
In addition, an equal-intensity configuration (blue curve) will maximize the peak intensity enhancement factor $\alpha = (\sum M_k)^2 / \sum M_k^2$ , in excellent agreement with the theoretical prediction in Sec.~\ref{sec:synthesis}. Hence the intensity ratio should be within the red and the blue curve to balance the peak intensity and the SNR of the synthesized APTs, which can be easily achieved through tuning the K of radiator segments. Furthermore, the linear phase-matching condition in Eq.~\eqref{phi} must be satisfied for the generation of synthesized APTs. We simulated the case where the relative phases of the three harmonics violate. And the result is shown in the magenta curve in Fig.~\ref{fig:power_syn_rad3_A_phix}. We can see the resulting waveform exhibits severe temporal degradation, in which a pronounced quasi-continuous power pedestal forms across the pulse trains, and the contrast between the main peak and adjacent sidelobes is drastically reduced. While in the blue curve in Fig.~\ref{fig:power_syn_rad3_A_phix}, the waveform synthesized under the matching condition maintains a clean temporal envelope with high peak-to-sidelobe contrast.

In conclusion, this work demonstrates a viable pathway for generating soft X-ray APTs with gigawatt peak power, sub-200 as pulse duration and highly waveform and spectrum tunability. Theoretical analysis and three-dimensional numerical simulation results have been presented to validate the proposed method. Detailed optimization of five harmonic synthesis gives out soft X-ray APTs with peak power of 3.5~GW and micropulse duration of 160~as. And the non-consecutive harmonic FEL lasing and synthesis, together with the shifts of the target harmonic components during the WS processes, demonstrate the flexible spectral and temporal control and adjustment of the synthesized APTs. 
The proposed method leverages precise control over harmonic components, relative phases, and radiator dispersions to shape the seeded FEL radiation pulses, thereby transforming the seeded FEL from a fully coherent single-harmonic source into a flexible platform for ultrafast laser pulse engineering. Beyond APTs' generation, the demonstrated capability to independently  manipulate the relative phases and amplitudes of multiple harmonics offers a powerful degree of freedom for ultrafast science. 
% Combining the WS technique with the high gain FEL principles may even further reach the zeptosecond-timescale pulse duration.  

%This approach not only enhances peak intensities and enables additional pulse compression but also provides a route toward generating isolated FEL pulses below 100~as through tailored spectral phase and amplitude profiles. Such sub-100-as soft X-ray pulses would open unprecedented opportunities for probing electron dynamics in atoms, molecules, and condensed matter on their natural timescales, potentially enabling real-time observation of processes such as Auger decay, charge migration, and valence-electron dynamics.
\section*{Acknowledgments}
%Anyone who made a contribution to the research or manuscript, but who is not a listed author, should be acknowledged (with their permission). Types of acknowledgements include:

% \subsection*{General} 
%Thank others for any contributions, whether it be direct technical help or indirect assistance. 

The authors thank Yaozong Xiao, Weijie Fan and Haiyang Li for helpful discussions and useful comments.

\subsection*{Author Contributions} 
%Describe contributions of each author to the paper, using the first initial and full last name. 

% \medskip Examples:

% ``S. Zhang conceived the idea and designed the experiments.''

% ``E. F. Mustermann and J. F. Smith conducted the experiments.''

% ``All authors contributed equally to the writing of the manuscript.''

Lanpeng Ni and Zheng Qi conceived the idea and scheme. Chao Feng and Zheng Qi supervised the project. Lanpeng Ni and Junhao Liu performed the simulations and analyzed the results. Contribution of all authors to the writing of the manuscript is equal.

\subsection*{Funding}
%Name financially supporting bodies (written out in full), followed by the funding awardee and associated grant numbers (if applicable) in square brackets. 

% \medskip Example: 

% ``This work was supported by the Engineering and Physical Sciences Research Council [grant numbers xxxx, yyyy]; the National Science Foundation [grant number zzzz]; and a Leverhulme Trust Research Project Grant.'' 

% \medskip
% If the research did not receive specific funding, but was performed as part of the employment of the authors, please name this employer. If the funder was involved in the manuscript writing, editing, approval, or decision to publish, please declare this.

This work was supported by the SXFEL facility, National Natural Science Foundation of China (Grant No.12435011, 12405363), Project for Young Scientists in Basic Research of Chinese Academy of Sciences (YSBR-115, YSBR-091).
\subsection*{Competing interests}
% Conflicts of interest (COIs, also known as ``competing interests'') occur when issues outside research could be reasonably perceived to affect the neutrality or objectivity of the work or its assessment. 

% Authors must declare all potential interests – whether or not they actually had an influence – in a ‘Conflicts of Interest’ section, which should explain why the interest may be a conflict. Authors must declare current or recent funding (including for Article Processing Charges) and other payments, goods or services that might influence the work. All funding, whether a conflict or not, must be declared in a ``Funding Statement.'' The involvement of anyone other than the authors who 1) has an interest in the outcome of the work; 2) is affiliated to an organization with such an interest; or 3) was employed or paid by a funder, in the commissioning, conception, planning, design, conduct, or analysis of the work, the preparation or editing of the manuscript, or the decision to publish must be declared.

% If there are none, the authors should state ``The author(s) declare(s) that there is no conflict of interest regarding the publication of this article.'' Submitting authors are responsible for coauthors declaring their interests. Declared conflicts of interest will be considered by the editor and reviewers and included in the published article.

The authors declare that there is no conflict of interest in the publishing this article.
\subsection*{Data Availability}
% A data availability statement is compulsory for all research articles. This statement describes whether and how others can access the data supporting the findings of the paper, including 1) what the nature of the data is, 2) where the data can be accessed, and 3) any restrictions on data access and why.

% If data are in an archive, include the accession number or a placeholder for it. Also include any materials that must be obtained through a Material Transfer Agreements (MTA). 
The corresponding author can provide relevant data if the demand is reasonable.

\printbibliography
\end{document}